\documentclass{ws-procs9x6}

\begin{document}

\title{Nuclear Model of Binding alpha-particles
\footnote{\uppercase{T}his work is partially supported by
\uppercase{D}eutsche \uppercase{F}orschungsgemeinschaft
   (grant 436 \uppercase{RUS} 113/24/0-4), \uppercase{R}ussian \uppercase{F}oundation
   for \uppercase{B}asic \uppercase{R}esearch
   (grant 03-02-04021) and the \uppercase{H}eisenberg-\uppercase{L}andau
   \uppercase{P}rogram (\uppercase{JINR})
}}

\author{K.A.~GRIDNEV, S.Yu.~TORILOV}

\address{Institute of Physics, St.Petersburg State University, 198904,
Russia}

\author{V.G.~KARTAVENKO}

\address{Joint Institute for Nuclear Research, Dubna, Moscow District, 141980,
Russia}

\author{W.~GREINER, D.K.~GRIDNEV}

\address{J.W.~Goethe University, Frankfurt/Main, D-60054, Germany}

\author{J.~HAMILTON}

\address{Vanderbilt University, Nashville, TN 37235, USA}

\maketitle

\abstracts{The model of binding alpha-particles in nuclei is
suggested.  It is shown good (with the accuracy of 1-2\%)
description of the experimental binding energies in light and
medium nuclear systems. Our preliminary calculations show
enhancement of the binding energy for super heavy nuclei with Z
$\sim$120.}

We assume that atomic nuclei might be considered as a tight
packing of alpha-particles. The binding energy could be expressed
as a sum of energies coming from interactions between $N_{\alpha}$
alpha-particles and their self-energies.
\begin{equation}\label{eb1}
    E_B(N_{\alpha}) = E_B(\alpha)N_{\alpha} + E_{C} + V_{\alpha\alpha}(N_{\alpha})
\end{equation}
where
$V_{\alpha\alpha}(N_{\alpha})=\sum_{j>i}^{N_{\alpha}}V_{\alpha\alpha}(r_{ij})$,
$E_B(\alpha)$=28.296~MeV is the binding energy of alpha-particle,
and $E_{C}$ is the Coulomb part of the potential energy.
For the alpha-alpha interaction  we use the Lennard-Jones (LJ)
potential, which can be written as
\begin{equation}\label{lj}
    V_{\alpha\alpha}(r) = V_{0} \left[ \left( \sigma/r \right)^{12}
    - 2 \left(\sigma/r \right)^{6} \right]
\end{equation}
\begin{figure}[htbp]
\begin{center}
\includegraphics[width=7.9cm]{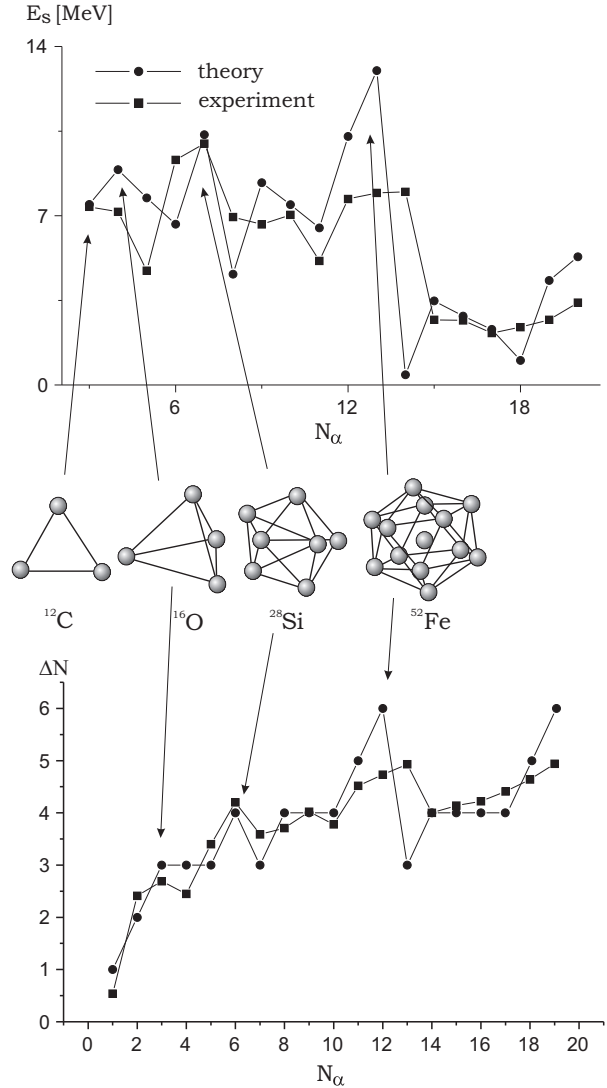}
\end{center} \caption{Separation energy of a single alpha-particle
versus the number of alpha-particles $N_{\alpha}$. The
alpha-particle sets and the differences of inter-alpha-particle
bonds ($\Delta N=B_{\alpha}(N_{\alpha}+1)-B_{\alpha}(N_{\alpha})$,
where $B_{\alpha}(N_{\alpha})$ is the number of the bonds for
nucleus with $N_{\alpha}$ alpha-particles.) \label{fig1}}
\vspace*{-1mm}
\end{figure}
where $r$ is the distance between centers of clusters, $V_{0}$ is
the depth of the potential ($V_{\alpha\alpha}(\sigma)=-V_{0},\quad
dV_{\alpha\alpha}/dr|_{r=\sigma}=0)$, and $\sigma$ is the pair
bonding length.
\begin{figure}[htbp]
\begin{center}
\includegraphics[width=10.5cm]{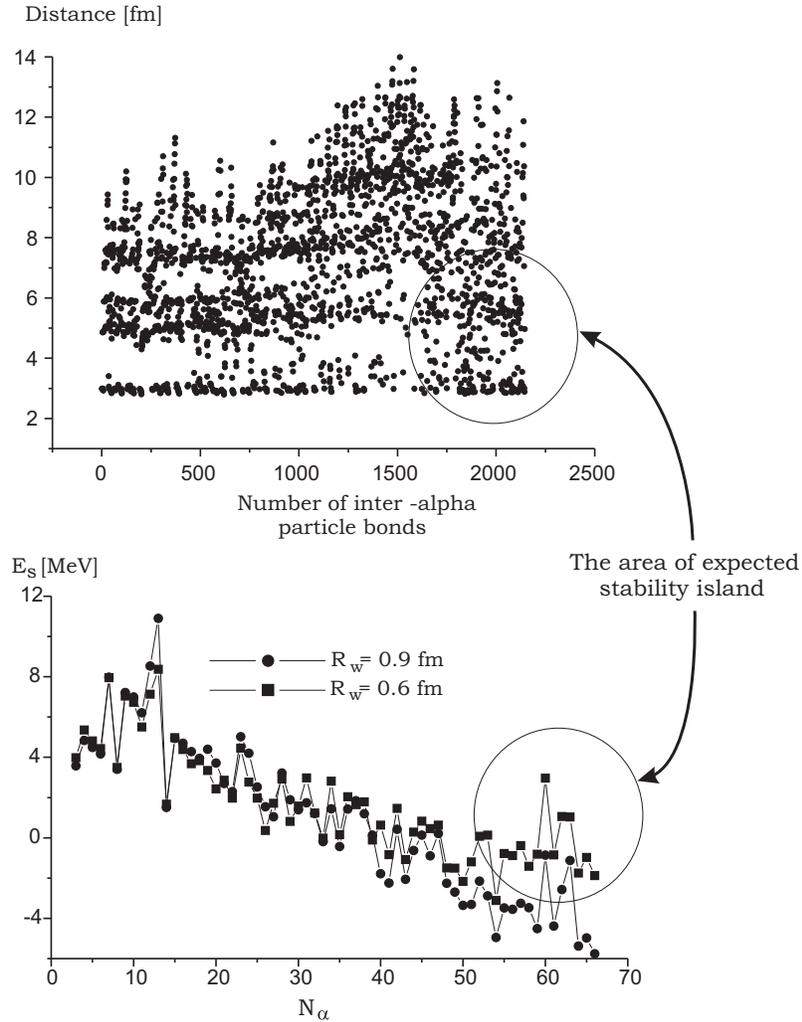}
\caption{The set of distances between alpha-particles in the
nucleus versus the number of bonds between alpha-particles
$B_{\alpha}$. Separation energy of an alpha-particle as the
function of the number of alpha-particles $N_{\alpha}$.
\label{fig2}}
\end{center}
\end{figure}
The positions of alpha-particles are found from the condition of
the minimal total potential energy.  The volume of the nucleus is
set on a grid and the first alpha-particle is positioned in the
center. The potential is calculated in all nodes of the grid and
the next particle is positioned into the node with the minimum
potential energy ($r=\sigma$). Then we take two particles and
calculate their potential and put to its minimum the third
cluster. And so on and so forth. This simple algorithm gives
possibility to cast the binding energy (\ref{eb1}) in the
following form\cite{norman}
\begin{equation}\label{eb2}
    E_B(N_{\alpha}) = E_B(\alpha) N_{\alpha} + V_0 B_{\alpha} + E'_{C}
\end{equation}
where the effective geometrical factor $B_{\alpha}$ is the number
of bonds between alpha particles, and all Coulomb terms are
collected in ${E'}_{C}$.  Assuming six equal nucleon  bonds in
each close-packed tetrahedral alpha particle, the unit bond energy
will approximately equal to
$V_0\approx~E_B(\alpha)/6\approx~$5~MeV.

The cluster structure of nuclear system, and the selected {\it
type of interaction} (LJ Eq.~(2) results to enhanced stability of
tight cluster configurations with {\it icosahedral symmetry}  (see
Figures).  It is  seen, in the upper part of Fig.~2, that the
first set of distances between alpha-particles are nearly
$\sigma\approx~3$~fm, and then there are stripes corresponding to
non-bordering alpha-particles and {\it etc}. Gradually the stripes
wash out because the tetrahedrons have to deform in order to
constitute closed figures.

The cluster structure of nuclear system leads to (many-body) shell
effects (see Figures), which are not reduced to the traditional
nucleon single-particle ones. Such shell effects could be very
important for heavy and superheavy nuclei, which stability is
defined mainly by the shell effects.

Assuming that the main trend remains when adding neutrons, our
calculations show enhancement of the binding energy for super
heavy nuclei with Z $\sim$120, which approximately corresponded to
the region of completed icosahedral shell of 55 alpha-particles.
One of us \cite{gr02sup} has predicted recently fullerene-like
nuclei in this region. The details (exact binding energy and the
$Z$ value) depend on the details of the potential, but the main
trends conserve, as seen from the lower part of Fig.~2, where
results are presented for two different values of the potential
width ($R_{W}=
2^{1/12}\sigma((\sqrt{2}+1)^{1/6}-(\sqrt{2}-1)^{1/6}$).

It turns out also, the structure of heavy even-even nuclei looks
like the Fe nucleus coated with the Bose condensate of
alpha-particles with additional neutrons playing the role of
electrons in the covalent bond. A similar mechanism of
alpha-particle formation analogous with formation of Cooper pairs
has been considered in \cite{koh}.


\begin{thebibliography}{0}
%
\bibitem{norman} P.D.~Norman, {\it Eur.~J.~Phys.} {\bf 14}, 36 (1993).
%
\bibitem{gr02sup}
W.Greiner, {\it Prog.~Theor.~Phys.~Suppl.} {\bf 146}, 84 (2002).
%
\bibitem{koh}
S.~Koh, {\it Prog.~Theor.~Phys.~Suppl.}  {\bf 132}, 197 (1998).
%
\end{thebibliography}
\end{document}